\documentclass{ws-ijmpd}
\begin{document}

\markboth{V. Van Elewyck {\it et al.}}
{Joint searches for gravitational waves and high-energy neutrinos}

\catchline{}{}{}{}{}
%
%%%%%%%%%%%%%%%%%%%%%%%%%%%%%%%%%%%%%%%%%%%%%%%%%%%%%%%%%%%%%%%%%%%%

\title{JOINT SEARCHES BETWEEN GRAVITATIONAL-WAVE INTERFEROMETERS AND HIGH-ENERGY NEUTRINO TELESCOPES: SCIENCE REACH AND ANALYSIS STRATEGIES}

\author{VERONIQUE VAN ELEWYCK$^{a}$\footnote{presenter; elewyck@apc.univ-paris7.fr} , \\
S.~ANDO$^{b}$, Y.~ASO$^{b}$, B.~BARET$^{a}$, M.~BARSUGLIA$^{a}$, I.~BARTOS$^{c}$, E.~CHASSANDE- MOTTIN$^{a}$, I.~DI~PALMA$^{c,d}$, J.~DWYER$^{c}$, C.~FINLEY$^{e}$, K.~KEI$^{f}$, A.~KOUCHNER$^{a}$, S.~MARKA$^{c}$, Z.~MARKA$^{c}$, J.~ROLLINS$^{c}$, C.~D.~OTT$^{b}$, T.~PRADIER$^{g}$, A.~SEARLE$^{b}$}
\address{(a) AstroParticule et Cosmologie (UMR 7164) \& Universit\'e Paris 7, \\Case 7020, F-75205 Paris Cedex 13, France \\
(b) California Institute of Technology, Mail Code 130-33, Pasadena, California 91125, USA \\
(c) Department of Physics, Columbia University, New York, NY, 10027, USA \\
(d) INFN \& University of Rome ”La Sapienza”, P.le Aldo Moro 2, 00185 Roma, Italy \\
(e) Department of Physics, University of Wisconsin, Madison,WI, 53706, USA \\
(f) Division of Theoretical Astronomy/Center for Computational Astrophysics, National Astronomical Observatory of Japan, 
2-21-1, Osawa, Mitaka, Tokyo, 181 8588, Japan \\
(g)Institut Pluridisciplinaire Hubert Curien \& Universit\'e Louis-Pasteur, Strasbourg (France)}
\maketitle

\begin{history}
\received{Day Month Year}
\revised{Day Month Year}
\comby{Managing Editor}
\end{history}

\begin{abstract}
Many of the astrophysical sources and violent phenomena observed in
our Universe are potential emitters of gravitational waves (GWs) and high-energy
neutrinos (HENs). A network of GW detectors such as LIGO and Virgo can determine the direction/time of GW bursts while the IceCube and ANTARES neutrino telescopes can also provide accurate directional information for HEN events. Requiring the consistency
between both, totally independent, detection channels shall enable new searches for cosmic events arriving from potential common sources, of which many extra-galactic objects.
\end{abstract}

\keywords{multi-messenger astronomy; neutrinos; gravitational waves}

\section{Introduction and motivations}	
\label{sec:intro}
Multi-messenger astronomy is entering an exciting period with the recent development of experimental techniques that have opened new windows of observation of the cosmic radiation in all its components. Gamma-ray astronomy - extensively discussed elsewhere in these Proceedings - has already shown evidence for (extra-)galactic sources of electromagnetic radiation with an energy spectrum extending up to several tens of TeVs; but most of them are expected to become fainter at higher energies due to the absorption of high-energy photons through interactions in the source and with the extragalactic background light. Many such sources are believed to originate from cataclysmic events associated with the production of gravitational waves (GWs), and with the emission of high-energy neutrinos (HENs) as a byproduct of the hadronic processes involving accelerated protons (and photons). Both HENs and GWs are alternative cosmic messengers that may escape very dense media and travel unaffected over cosmological distances, carrying information on the internal processes of the astrophysical engines.

Joint GW-HEN searches are also motivated by the advent of a new generation of dedicated detectors. The first undersea neutrino telescope, ANTARES, is now operating in its final 12-line (12L) configuration, covering an instrumented volume of about 0,04 km$^3$ in the Mediterranean Sea\cite{antares}, while at the South Pole, IceCube has been taking data with an increasing number of lines (59 by now) and is proceeding towards its full, km$^3$-sized, configuration\cite{ice3}. The field of view of these detectors is about $2 \pi$ sr for neutrino energies $100\ \mathrm{GeV} \leq E_\nu \leq 100\  \mathrm{TeV}$, with an instantaneous overlap of $\sim 0.5\ \pi$ sr. Both have good directional resolution, with a median error between $1^\circ$ and $2^\circ$ for  IceCube and possibly as small as $\sim 0.3^\circ$ above 10 TeV for ANTARES as a result of the optical properties of sea water. 

The GW detectors VIRGO\cite{virgo} (with one site in Italy) and LIGO\cite{ligo} (with two sites in the United States) are Michelson-type laser interferometers that consist of two light storage arms oriented at $90^\circ$ from each other, with suspended mirrors playing the role of test masses; their current detection horizon is about 15 Mpc for standard binary sources. 
Both detectors had a data-taking phase during 2007, which partially coincided with the ANTARES 5L and IceCube 22L configurations. They are currently upgrading to improve their sensitivity by a factor of 2 - and hence the probed volume by a factor of 8 - and are preparing for a common science run starting mid-2009, i.e. in coincidence with the operation of ANTARES 12L and IceCube 59L. The VIRGO/LIGO network monitors a good fraction of the sky in common with HEN telescopes: the overlap of visibility maps with each telescope is about 4~sr ($\sim 30\%$ of the sky).

\section{Potential common sources of GW and HEN}
\label{sec:sources}

Potential sources of GWs and HENs are likely to be very energetic and to exhibit bursting activity. 
Plausible GW+HEN emission mechanisms include two classes of galactic sources which could be accessible to the present generation of GW interferometers and HEN telescopes. {\bf Microquasars} are believed to be X-ray binaries involving a compact object that accretes matter from a companion star and re-emits it in relativistic jets associated with intense radio (and IR) flares. Such objects could emit GWs during both accretion and ejection phases; and the latter phase could be correlated with a HEN signal as well if the jet has a hadronic component\cite{migliari}. {\bf Soft Gamma Repeaters (SGRs)} are X-ray pulsars with a soft $\gamma$-ray bursting activity which, according to the magnetar model, can be associated with  star-quakes. The deformation of the star during the outburst could produce GWs, while HENs could emerge from hadron-loaded flares\cite{ioka}. 

Gamma-Ray Bursts (GRBs) are another promising class of extragalactic sources. In the prompt and afterglow phases, HENs ($10^5 - 10^{10}$ GeV) are expected to be produced by accelerated protons in relativistic shocks and several models predict detectable fluxes in km$^3$-scale detectors\cite{GRBnus}.{\bf Short-hard GRBs} are thought to originate from coalescing binaries involving  black holes and/or neutron stars; such mergers could emit GWs detectable from relatively large distances, with significant associated HEN fluxes\cite{nakar}. As for the {\bf long-soft GRBs}, the collapsar model is compatible with the emission of a strong burst of GWs during the gravitational collapse of the (rapidly rotating) progenitor star and in the pre-GRB phase; however this population is distributed over cosmological distances so that the associated HEN signal is expected to be faint\cite{kotake}. The subclass of {\bf low-luminosity GRBs}, with $\gamma$-ray luminosities a few orders of magnitude smaller, are believed to originate from a particularly energetic, possibly rapidly-rotating and jet-driven population of core-collapse supernovae. They could produce stronger GW signals together with significant high- and low-energy neutrino emission; moreover they are often discovered at shorter distances\cite{gupta}. Finally, the {\bf failed GRBs} are thought to be associated with supernovae driven by mildly relativistic, baryon-rich and optically thick jets, so that no $\gamma$-rays escape. Such ``hidden sources'' could be among the most promising emitters of GWs and HENs, as current estimations predict a relatively high occurrence rate in the volume probed by current GW and HEN detectors\cite{ando}.

\section{Outlook on the analysis strategies}
\label{sec:ana}

GW interferometers and HEN telescopes share the challenge to look for faint and rare signals buried in abundant noise or background events. The GW+HEN search methodology involves the combination of independent GW/HEN candidate event lists, with the advantage of significantly lowering the rate of accidental coincidences. 

The information required about any GW/HEN event consists of its timing, arrival direction and associated angular uncertainties (possibly under the form of a sigificance sky map). Each event list is obtained by the combination of  reconstruction algorithms specific to each experiment, and quality cuts used to reject as much background as possible.
GW+HEN event pairs within a predefined, astrophysically motivated (and possibly source- or model-dependent), time interval can be selected as time-coincident events. Then, the spatial overlap between GW and HEN events is statistically evaluated, e. g. by an unbinned maximum likelihood method, and the significance of the coincident event is obtained by comparing to the distribution of accidental events obtained with Monte-Carlo simulations using data streams scrambled in time (or simulated background events).

Preliminary investigations of the feasibility of such searches have already been performed and indicate that, even if the constituent observatories provide several triggers a day, the false alarm rate for the combined detector network can be maintained at a very low level ($\sim (600\ \mathrm{yr})^{-1}$)\cite{aso,pradierVLVNT}.

\section{Conclusions and perspectives}
\label{sec:concl}

A joint GW+HEN analysis program could significantly expand the scientific reach of both GW interferometers and HEN telescopes. The robust background rejection arising from the combination of two totally independent sets of data results in an increased sensitivity and the possible recovery of cosmic signals. The observation of coincident triggers would provide strong evidence for the detection of a GW burst and a cosmic neutrino event, and for the existence of common sources. Beyond the benefit of a high-confidence discovery, coincident GW/HEN (non-)ob\-servation shall play a critical role in our understanding of the most energetic sources of cosmic radiation and in constraining existing models. They could also reveal new, ``hidden'' sources unobserved so far by conventional photon astronomy. 
A new period of concurrent observations with upgraded experiments is expected to start mid-2009. Future schedules involving next-generation detectors with a significantly increased sensitivity (such as the future km$^3$-sized undersea neutrino telescope, KM3NeT\cite{km3net}, and the Advanced LIGO/Advanced VIRGO projects) are likely to coincide as well, opening the way towards an even more efficient GW+HEN astronomy.
\section*{Acknowledgments}

V. V. E.  warmly thanks the organizers for the lively and friendly atmosphere of the workshop. The authors acknowledge financial support from the EC 7th Framework Program (Marie Curie Reintegration Grant), the NSF (grants PHY-0107417 /0757058/0457528/0757982) and the ANR (contract ANR-08-JCJC-0061-01).

%\begin{thebibliography}{000} %for 3 digits
%\begin{thebibliography}{00}  %for 2 digits

\end{document}